\documentclass[preprint]{aastex631}
\usepackage{bm, amsmath, booktabs, multirow, natbib}
  \def\kms{{\rm km}\,{\rm s}^{-1}} \def\masyr{{\rm mas}\,{\rm yr}^{-1}}   \def\mas{{\rm mas}}       \def\max{{\rm max}}  \def\rel{{\rm rel}}  \def\eff{{\rm eff}}     \def\B{{\rm b}} \def\Sc{{\rm s}} \def\L{{\rm l}} \def\E{{\rm E}}            

\begin{document}

\title{``Industrial-Scale'' Black Hole Selection without a Satellite}

\correspondingauthor{Subo Dong}
\email{dongsubo@pku.edu.cn}

\author[0009-0007-5754-6206]{Zexuan Wu}
\affil{Department of Astronomy, School of Physics, Peking University,
	5 Yiheyuan Road, Haidian District, Beijing 100871, People's Republic of China \\}
\affil{Kavli Institute of Astronomy and Astrophysics, Peking University,
	5 Yiheyuan Road, Haidian District, Beijing 100871, People's Republic of China \\}
\author[0000-0002-1027-0990]{Subo Dong}
\affil{Department of Astronomy, School of Physics, Peking University,
	5 Yiheyuan Road, Haidian District, Beijing 100871, People's Republic of China \\}
\affil{Kavli Institute of Astronomy and Astrophysics, Peking University,
	5 Yiheyuan Road, Haidian District, Beijing 100871, People's Republic of China \\}
\affil{National Astronomical Observatories, Chinese Academy of Science, 20A Datun Road, Chaoyang District, Beijing 100101, People's Republic of China\\}
\author{Andrew P. Gould}
\affil{Max Planck Institute for Astronomy, K\"{o}nigstuhl 17, 69117 Heidelberg, Germany\\}
\affil{Department of Astronomy, Ohio State University, 140 W. 18th Ave., Columbus, OH 43210, USA\\}
\author[0000-0001-7016-1692]{Przemek Mr{\'o}z}
\affil{Astronomical Observatory, University of Warsaw, Al. Ujazdowskie 4, 00-478 Warszawa, Poland\\}
\author[0000-0003-2125-0183]{A.~M\'erand}
\affil{European Southern Observatory, Karl-Schwarzschild-Stra{\ss}e 2, D-85748 Garching, Germany\\}

\begin{abstract}

The forthcoming GRAVITY+ instrument promises to usher in an era of ``industrial-scale''  mass measurements of isolated black holes (BHs), with the potential to assemble a sample of many tens of BHs via interferometric microlensing over several years. A key challenge will be selecting interferometric follow-up targets from an order-of-magnitude larger pool of ongoing microlensing events when using traditional selection based on event timescale alone. \citet{Gould23} proposed a criterion optimized for BH selection using space-based microlens parallax measurements enabled by a satellite. We adapt it to work with microlens parallax constraints obtainable from ground-based data only. Using Galactic simulations, we show that our selection criterion is highly efficient, expecting to detect about a dozen BHs per year with GRAVITY+ from following up $\sim35$ selected events.

\end{abstract}


\section{Introduction} \label{sec:intro}
Stellar-mass black holes (BHs) can be produced at the end of massive stellar evolution. 
The Galaxy may host $\sim 10^8\text{--}10^9$ stellar-mass BHs \citep{sat83, vandenheuvel92, bab94, Timmes96}, but few are observationally confirmed due to the difficulty of detection.

Known stellar-mass BHs are almost exclusively identified in binary systems. The Galactic systems include two dozen BHs detected through the long-practiced method of X-ray binaries \citep{ram06, BlackCAT} and a few BHs in wide binaries via recent radial velocity and the {\it Gaia} astrometric surveys \citep[see review by][]{Gaiabinary}. In addition, over a hundred extragalactic BH systems have been discovered in merging compact binaries via gravitational-wave radiation by LIGO \citep{LIGO23, LIGO25}.

Isolated  (i.e., single and non-accreting) BHs are expected to be abundant in the Galaxy, comprising a significant fraction of the overall stellar-mass BH population \citep[see, e.g.,][]{Wiktorowicz19, Olejak20}. 
However, only one isolated BH has been confirmed to date: OGLE-2011-BLG-0462, identified via astrometric microlensing using the {\it Hubble Space Telescope} ({\it HST}) data collected over a decade \citep{Sahu22, Mroz22, Lam23, Sahu25}.

Microlensing is the only known method for detecting isolated, dark objects such as stellar-mass BHs. In a microlensing event \citep{Paczynski86}, a foreground object (the ``lens'') temporarily magnifies the light of a background star (the ``source''), producing two images near the angular Einstein radius \citep{Einstein36}:
\begin{equation}
	\theta_\E = \sqrt{\kappa M \pi_\rel};\quad \kappa = \frac{4G}{c^2\rm au} \simeq 8.14 {\frac{\mas}{M_\odot}};\quad \pi_\rel = \pi_\L - \pi_\Sc,
\end{equation}
where $M$ is the lens mass, $\kappa$ is a constant, and $\pi_\rel$ is the relative lens-source 
trigonometric parallax. The relative proper motion between the lens and source, ${\bm \mu_\rel}$, induces a transient brightening event with a characteristic timescale (the Einstein timescale) $t_\E = \theta_\E / \mu_\rel$, which is routinely measured from a microlensing light curve. 
All else being equal, $t_\mathrm{E} \propto \sqrt{M}$, so a BH lens, being more massive than a typical stellar lens, tends to produce a longer $t_\mathrm{E}$. 
However, the timescale $t_\E$ is a degenerate combination involving the lens mass, distance, and transverse velocity. 

The central challenge in unambiguously identifying a BH lens lies in measuring its mass \citep{Gould00, Agol02}.  To break the parameter degeneracy and determine the mass of a dark lens, one must measure two additional quantities: the Einstein radius $\theta_\E$ and the microlens parallax vector $\bm{\pi}_\E$ \citep{Gould04}:
\begin{equation}
	\bm \pi_\E = \frac{\pi_\rel}{\theta_\E}\frac{\bm \mu_\rel}{\mu_\rel},
\end{equation}
With both $\theta_\E$ and $\bm{\pi}_\E$ measured, the degeneracy in determining the physical properties of the lens can be fully lifted:
\begin{equation}
	M=\frac{\theta_{\mathrm{E}}}{\kappa \pi_{\mathrm{E}}} ; \quad \pi_{\mathrm{rel}}=\theta_{\mathrm{E}} \pi_{\mathrm{E}} ; \quad \boldsymbol{\mu}_{\mathrm{rel}}=\frac{\theta_{\mathrm{E}}}{t_{\mathrm{E}}} \frac{\boldsymbol{\pi}_{\mathrm{E}}}{\pi_{\mathrm{E}}}
\end{equation}
The microlens parallax $\bm{\pi}_\E$ can be constrained either from simultaneous observations at two widely-separated sites \citep{Refsdal66} or from light-curve distortions due to Earth's orbital motion \citep{Gould92}. Measuring $\theta_\E$, however, is the principle obstacle for detecting BH lenses \citep[see relevant discussions in][]{Dong19, Gould23BD}. Although {\it HST} astrometric microlensing has delivered a breakthrough by measuring $\theta_\E$ of OGLE-2011-BLG-0462, the approach is too slow (one in a decade) to accumulate a statistically significant sample.

A direct and precise measurement of $\theta_\E$ can be achieved with long baseline ($\gtrsim 100$\,m) optical/infrared interferometry, whose milliarcsecond (mas) angular resolution can resolve the two microlensed images separated by $\simeq 2\theta_\E$ \citep{Delplancke01, Dalal03, Rattenbury06, Cassan16}. Unlike astrometric microlensing, which requires multi-epoch data usually spanning years to measure $\theta_\E$, the interferometric microlensing needs only one (or possibly two, closely-spaced) epochs. 
The GRAVITY instrument \citep{gravity17} on the Very Large Telescope Interferometer (VLTI) enabled the first resolution of microlensed images \citep{Dong19}. However, the \citet{Dong19} detection was near VLTI/GRAVITY's sensitivity limit, restricting the observable events to only a handful of exceptionally bright ones per year.

The ongoing GRAVITY+ upgrade marks a transformative sensitivity enhancement \citep{gravityplus, vltigpao}.  Recently, \citet{Mroz25} reported the first microlensing observation with GRAVITY Wide \citep{gravitywide}, an intermediate phase of the upgrade, demonstrating its greatly boosted capability to access dozens of long-timescale microlensing events ($t_{\rm E}>50$\,d) per year. Once fully operational, GRAVITY+ is expected to be further enhanced, capable of detecting $\sim10\text{--}20$ BHs per year, thereby enabling the rapid collection of a large sample.

However, such ``industrial-scale'' BH production raises a challenge of target selection. Only selecting long events with $t_{\rm E}>50$\,d would require VLTI to follow up an order of magnitude more non-BH events, potentially placing heavy demand on resources. To address this, \citet{Gould23} proposed an optimized strategy that additionally vets events using a $\pi_\E/t_{\rm E}$ criterion, with $\pi_{\rm E}$ measured from a satellite.

Motivated by \citet{Gould23}, we develop a BH event selection criterion relying on available ground-based data, i.e., without a satellite, and evaluate its performance with Galactic simulations.

\section{Selection Criteria} \label{sec:sel}

\citet{Gould23} introduced the following $\pi_\E/t_{\rm E}$ criterion for selecting VLTI follow-up events:
\begin{equation}
	\label{eq:mu_thresh}
	\pi_\E < \frac{\mu_{\rm thresh} t_\E}{\kappa M_\odot} = \frac{t_\E}{8.14\,\rm yr}\frac{\mu_{\rm thresh}}{1\,\masyr},
\end{equation}
where $\mu_{\rm thresh}$ is a constant expressed in units of proper motion. This criterion is equivalent to a mass-dependent threshold for relative proper motion of $\mu_{\rm rel}<\mu_{\rm thresh} (M/M_\odot)$. As shown by \citet{Gould23}, a reasonable value is $\mu_{\rm thresh}=1.5$\,mas/yr, corresponding to  $\pi_\E < 0.025\,(t_{\rm E}/50\,{\rm d})$.  Applying it effectively eliminates a significant fraction of low-mass ``contaminants'' ($M\lesssim 1\,M_\odot$) while retaining most BHs. See \S~2.2.2 of \citet{Gould23} for a detailed discussion.  

From ground-based-only light curves, the microlens parallax vector $\bm{\pi}_\E$ is often poorly constrained at the time of interferometric follow-up. Instead, the so-called one-dimensional (1D) microlens parallax is typically much better constrained, namely the component of $\bm{\pi}_\E$ parallel to Earth's projected acceleration, $\pi_{\E,\parallel}$ \citep{Gould94_1dplx, Smith03, Gould04}. 

We modify the $\pi_\E/t_{\rm E}$ criterion by replacing $\pi_\E$ in Equation~\ref{eq:mu_thresh} with $\pi_{\E,\parallel}$. Throughout this paper, we evaluate the following $\pi_{\E,\parallel}/t_{\rm E}$ criterion,
\begin{equation}
	\label{eq:piepar}
	 \pi_{\E,\parallel} < 0.025\,(t_{\rm E}/50\,{\rm d}).
\end{equation}

\section{Galactic simulations} \label{sec:gal}

We adopt simple toy models and perform simulations of microlensing events to assess the effectiveness of this VLTI selection criterion.  Our Galactic simulations mostly follow \citet{Jung21}, an updated implementation of the \citet{hangould03} model widely used in microlensing studies, with the main modification being the treatment of black hole lenses. 

\subsection{BH recipe}

Our basic approach to assigning remnant masses follows \citet{Gould00}, in which ascending mass thresholds in the stellar initial mass function (\citealt{Chabrier03} in our case) determine whether progenitors become white dwarfs (WDs), neutron stars (NSs), or BHs. While $8\,M_\odot$ is generally regarded as the boundary between WDs and NSs, the initial-final mass relation mapping progenitor masses to remnant masses of BHs and NSs is strongly debated \citep[see, e.g., review by][]{Pejcha20}. Given this uncertainty, we adopt a simple cut at $20\,M_\odot$: progenitors with $8 < M/M_\odot \leq 20\,M_\odot$ form NSs, and those with $M > 20\,M_\odot$ become BHs.

For the BH mass function (BHMF) in the simulations, we adopt an empirical approach. BHs in X-ray binaries are found to cluster around $\sim 7\text{--}8\,M_\odot$ \citep{Bailyn98, Ozel10}, consistent with the isolated BH OGLE-2011-BLG-0462 at $\sim 7.9\,M_\odot$ \citep{Mroz22}. In comparison, the LIGO binary BH mass distribution exhibits two overdensities: one near $\sim 10\,M_\odot$, similar to the peak of X-ray binaries, and another at $\sim 35\,M_\odot$  \citep{gwtc3, IAS25mass}. Interestingly, the three Gaia astrometric BHs fall near these LIGO peaks: two at $\sim 9\text{--}10\,M_\odot$ \citep{GaiaBH1, GaiaBH2} close to the lower peak, and Gaia-BH3 at $\sim 33\,M_\odot$ near the higher peak \citep{GaiaBH3}. Motivated by these observations, we consider the following two BHMFs in our simulations:

\begin{enumerate}
	\item {{\tt GWTC-3}:} A standard $\tt{POWER\ LAW}+\tt{PEAK}$ mass function, which is a truncated power-law distribution with a smooth low-mass turn-on and a high-mass peak, inferred from the Gravitational-Wave Transient Catalog 3 (GWTC-3) \citep{gwtc3}.
% \ ({\tt PP})
	\item {{\tt {\"O}zel+10}:} A Gaussian distribution $(7.8 \pm 1.2\, M_\odot)$ derived from BHs in X-ray binary systems \citep{Ozel10}.
\end{enumerate}

\begin{figure}
	\centering
\includegraphics[width=0.5\linewidth]{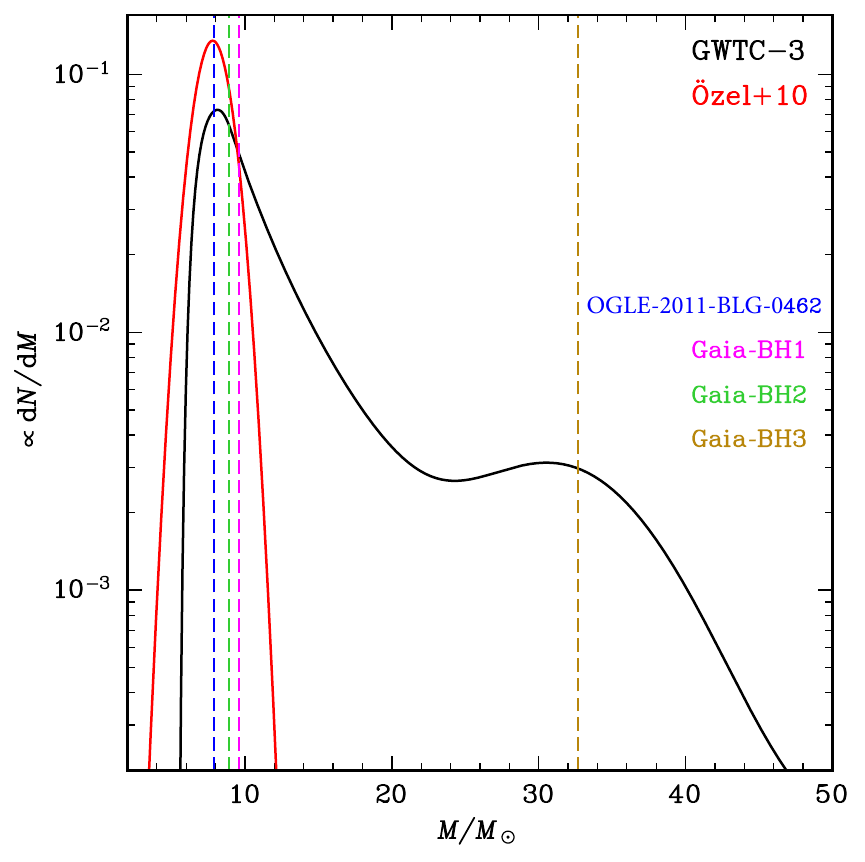}
	\caption{Black hole mass functions: {\tt{GWTC-3}} (black), based on LIGO BHs, and {\tt{{\"O}zel+10}} (red), inferred from X-ray binaries. Vertical dashed lines mark the mass estimates of the isolated BH OGLE-2011-BLG-0462 (blue) and three Gaia BHs (magenta, green, brown) in wide binaries.}
	\label{fig:BHMF}
\end{figure}

Figure~\ref{fig:BHMF} shows these BHMFs, along with the BH mass estimates of the individual systems mentioned above.

\subsection{Natal kicks}
\label{sec:kicks}

The high space velocities of neutron stars are generally thought to arise from ``natal kicks'' imparted by asymmetric supernova explosions at their birth \citep[see, e.g.,][and references therein]{Lai01}. In our simulations, we adopt a Maxwellian velocity distribution for NSs \citep{Hobbs05}, with a dispersion of $\sigma=217\,\text{km}\,\text{s}^{-1}$ \citep{Disberg25}. Whether and to what extent BHs receive natal kicks is, however, far less certain. Some studies on the spatial and velocity distributions of X-ray binaries suggest some BHs may receive substantial kicks of $\sim100\,\text{km}\,\text{s}^{-1}$ \citep[e.g.,][]{Repetto17, Zhao23}, whereas analyses of the isolated BH and certain cases of X-ray and wide binaries indicate much lower, or even negligible, kicks \citep{Sahu22, Andrews22, Koshimoto24, Burdge24, GaiaBH1, GaiaBH2}.

Owing to these uncertainties, we assume that the velocity distribution of BHs follows that of their birth stellar populations (i.e., zero kick) in our default analysis. 

We also explore simple Maxwellian distributions for BH kicks with mean velocities $\bar{v}_{\rm kick}$ of $100$, $200$, and $400\,\text{km}\,\text{s}^{-1}$. Natal kicks can increase averaged $\mu_{\rm rel}$, leading to shorter mean $t_\E$. In addition, they alter the spatial distributions of the lenses, most notably increasing the disk scale height, which in turn reduces the event rate \citep[see, e.g.,][]{Sweeney24}. We employ simplified treatments in our toy models. For the disk, we follow \citet{Koshimoto24} by using their analytical models fitted to numerical integration results of \citet{Tsuna18}. A weight factor is applied to the event rate to account for the modified Galactic scale height and surface density, which depend on the kick velocity, Galactocentric distance and vertical velocity dispersion. For the bulge, we conduct simple numerical experiments on BH orbital evolution with varying kick velocities in the Galactic potential, using the \texttt{galpy} package \citep{galpy}. The BHs are initially distributed according to an exponential density profile (effective radius $r_{e} = 1$\,kpc) with a 1D velocity dispersion of $\sigma = 100\,\kms$. After reaching a steady-state distribution, we estimate the best-fit effective radius as a function of $\bar{v}_{\rm kick}$, with $r_{e} = 1.10, 1.46, \text{and}\  2.36$\,kpc for $\bar{v}_{\rm kick}=100, 200, \text{and}\ 400\,\text{km}\,\text{s}^{-1}$, respectively, and apply the corresponding weight factor to the event rate.

\section{Results}
Our Galactic simulations yield catalogs of microlensing events, from which we generate mock light curves with $1$-day cadence and photometric uncertainties based on OGLE-IV data \citep{OGLE4error}. The peak time $t_0$ and impact parameter $u_0$ are randomly drawn from uniform distributions during the bulge observing season and between 0 and 1, respectively. In practice, $\pi_{\E,\parallel}$ will be measured as late as possible prior to the VLTI observing window in order to make the most accurate measurement possible. In this paper, we represent this decision time by $t_{\rm VLTI} = t_0 + \max\{t_\eff, 5\,\text{d}\}$, where $t_\eff\equiv t_{\rm E}u_0$ is the effective timescale, which is often well determined from incomplete light curves \citep{Yoo04, Dong06}.  We derive $\pi_{\E,\parallel}$ by fitting the mock light curve up to $t_{\rm VLTI}$.

\subsection{Default Yields}
\begin{table}[h!]
	\caption{Annual yields of $t_{\mathrm{E}} > 50\,\rm d$ events from the default simulations.}
	\label{tab:expected_yields}
	\centering
	\begin{tabular}{llrrrrrr}
		\toprule
		$\pi_\E$ vetting                            & BHMF           & $f_{\rm BH}$ & $n_{\rm BH}$ & $n_{\rm NS}$ & $n_{\rm WD}$ & $n_{\rm MS}$ & $n_{\rm sel}$ \\

		\midrule
		\multirow{2}{*}{none}                       & \tt{GWTC-3}    & 15.3\%       & 15.3         & 1.2          & 20.7         & 62.8         & 100.0         \\
		                                            & \tt{\"Ozel+10} & 11.4\%       & 11.4         & 1.3          & 21.6         & 65.7         & 100.0         \\
		\midrule
		\multirow{2}{*}{$\pi_\E/t_\E$}              & \tt{GWTC-3}    & 87.2\%       & 15.0         & 0.1          & 0.6          & 1.4          & 17.2          \\
		                                            & \tt{\"Ozel+10} & 82.4\%       & 10.8         & 0.1          & 0.7          & 1.5          & 13.1          \\
		\midrule
		\multirow{2}{*}{$\pi_{\E, \parallel}/t_\E$} & \tt{GWTC-3}    & 42.8\%       & 15.2         & 0.5          & 5.4          & 14.5         & 35.7          \\
		                                            & \tt{\"Ozel+10} & 34.6\%       & 11.3         & 0.5          & 5.6          & 15.2         & 32.7          \\
		\bottomrule
	\end{tabular}
\end{table}

\begin{figure*}
	\centering
	\begin{minipage}{1.0\linewidth}
		\makebox[0.5\linewidth]{\includegraphics[width=0.45\linewidth]{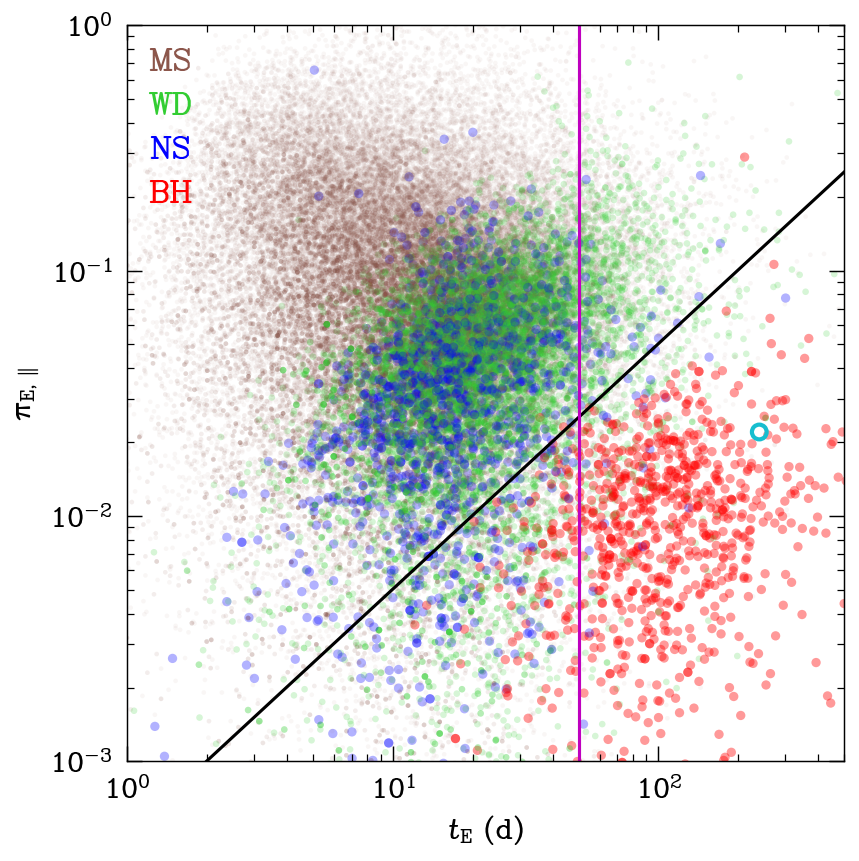}}%
		\makebox[.5\linewidth]{\includegraphics[width=0.45\linewidth]{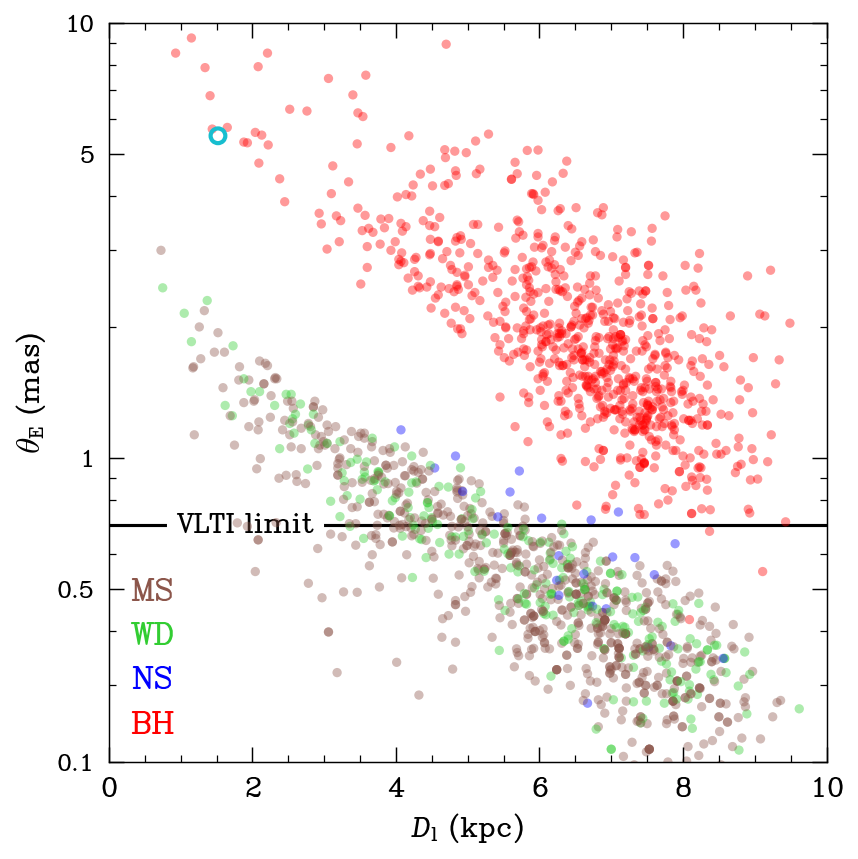}}
	\end{minipage}
    \caption{Left: Distribution of timescale $t_\E$ vs. 1D microlens parallax $\pi_{\E,\parallel}$ for mock microlensing events from the default (no-kick) simulations using the {\tt GWTC-3} BHMF. Events are color-coded by lens type: brown dwarfs and main-sequence stars (MS) in brown, white dwarfs (WD) in green, neutron stars (NS) in blue, and black holes (BH) in red. The boundaries of the two VLTI follow-up selection criteria, $t_\E > 50$\,d and $\pi_{\E,\parallel} < 0.025\,(t_{\rm E}/50\,{\rm d})$, are marked by a vertical magenta line and a sloped black line, respectively, which effectively separate the BHs from most of the contaminants. Right: Distribution of lens distance $D_\L$ vs. angular Einstein radius $\theta_{\E}$ of all the selected events. The horizontal line marks the $\theta_\E > 0.7$\,mas resolvability limit of VLTI, above which most of the BHs are included. In both panels, the confirmed isolated BH OGLE-2011-BLG-0462 is marked with a cyan open circle.}
	\label{fig:tE_piEparallel}
\end{figure*}
	
Table~\ref{tab:expected_yields} summarizes the expected annual yields of BHs and contaminants from the default (no-kick) simulations. By analyzing OGLE-IV data, we find that GRAVITY+ \citep{gravityplus, vltigpao} is capable of observing approximately 100 long events with $t_\E > 50\,\text{d}$ per year. We then simply normalize the total annual number of simulated $t_\E > 50\,\text{d}$ events to $100$, noting that for such long timescales the event detection efficiency in surveys is virtually independent of $t_\E$. Among these events, $f_{\rm BH}=15.3\%\ (11.4\%)$ are BHs for the {\tt GWTC-3} ({\tt {\"O}zel+10}) mass function, consistent with the $\sim10-20\%$ estimates from simulations by \citet{Gould00} and \citet{Lam20}, despite their different recipes for BH distributions. 

As shown in Table~\ref{tab:expected_yields}, the $\pi_{\E,\parallel}/t_{\rm E}$ criterion is highly effective in filtering out populations of brown dwarfs and main-sequence stars (collectively referred to as MS throughout the paper), as well as white dwarfs (WD) and neutron stars (NS). Out of 100 events with $t_\E > 50\,\text{d}$ per year, the total number of selected events that pass the $\pi_{\E,\parallel}/t_{\rm E}$ criterion is $n_{\rm sel}=35.7\ (32.7)$, with $f_{\rm BH}=42.8\%\ (34.6\%)$ of them being BHs. Importantly, only a negligible fraction of BHs are excluded by the criterion. Notably, if $\pi_{\E}$ can be constrained using a satellite, as envisioned by \citet{Gould23}, the $\pi_{\E}/t_{\rm E}$ criterion is extremely effective, among the $n_{\rm sel}\sim15$ events that pass vetting, $f_{\rm BH}\sim 80-90\%$ of them are BHs.

The left panel of Figure~\ref{fig:tE_piEparallel} shows that BHs are effectively separated from other lens populations in the $t_{\rm E}$--$\pi_{\E,\parallel}$ plane, illustrating the effectiveness of our vetting criteria. For the selected events, most of the BHs, except for a small fraction of low-mass BHs in the bulge, have $\theta_\E$ larger than $0.7$\,mas, which is the limit that can be resolved by VLTI (see the right panel of Figure~\ref{fig:tE_piEparallel}). Successful sub-percent level $\theta_\E$ measurements have been achieved with VLTI near this limit by \citet{Cassan22} with $\theta_\E = 0.765 \pm 0.004$\,mas and by \citet{Wu24} with $\theta_\E = 0.724 \pm 0.002$\,mas. 
 
\subsection{Effects of Natal Kicks}

For simulations with non-zero $\bar{v}_{\rm kick}$, we calculate $\eta_{\rm BH}$, the number of BHs among events that satisfy our selection criteria ($t_{\rm E}>50$\,d and $\pi_\E < 0.025\,(t_{\rm E}/50\,{\rm d})$), relative to the default (zero-kick) simulations, with results presented in Table~\ref{tab:relnum}.

\begin{table}[h!]
	\centering
	\caption{Relative number of vetted BHs $(\eta_{\rm BH})$ with various mean kick velocities $\bar{v}_{\rm kick}$ to the default (no-kick).
	}

	\label{tab:relnum}
	\begin{tabular}{lcc}
		\toprule
		BHMF               & \textbf{$\bar{v}_{\rm kick}$} $(\kms)$ & $\eta_{\rm BH}$ \\
		\midrule
		\multirow{4}{*}{\tt{GWTC-3}}      & 0                             & 1.000                  \\
		                             & 100                           & 0.901                  \\
		                             & 200                           & 0.578                  \\
		                             & 400                           & 0.211                  \\
		\midrule
		\multirow{4}{*}{\tt{\"Ozel+10}} & 0                             & 1.000                  \\
		                             & 100                           & 0.864                  \\
		                             & 200                           & 0.501                  \\
		                             & 400                           & 0.136                  \\
		\bottomrule
	\end{tabular}
\end{table}

\begin{figure*}
	\centering
	\begin{minipage}{1.0\linewidth}
		\makebox[.5\linewidth]{\includegraphics[width=.45\linewidth]{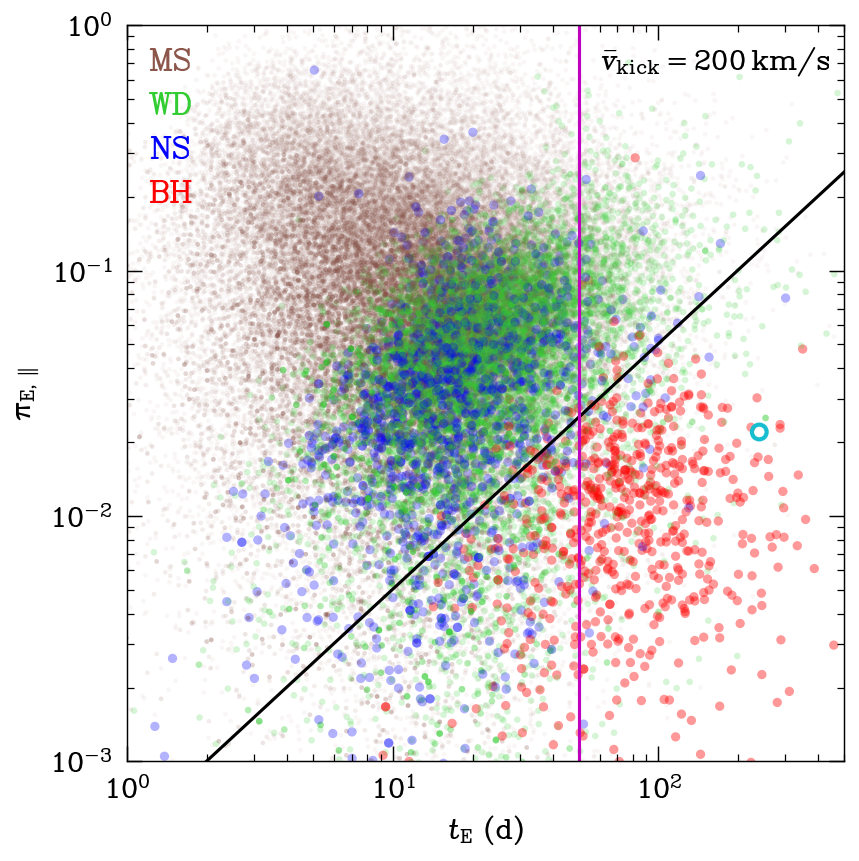}}%
		\makebox[.5\linewidth]{\includegraphics[width=.45\linewidth]{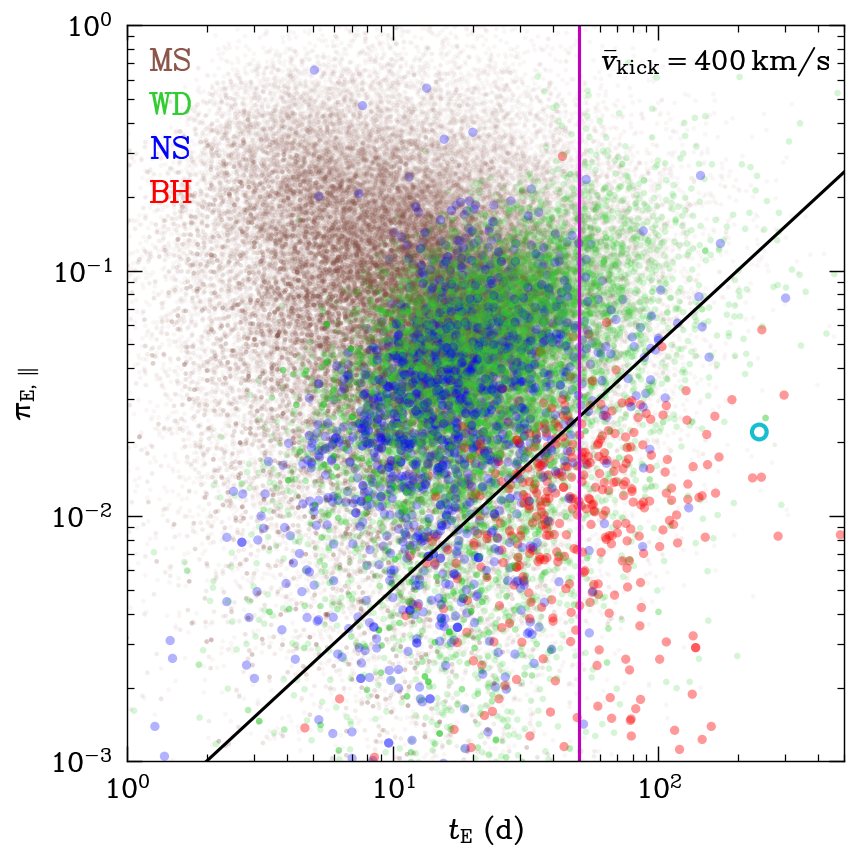}}%
	\end{minipage}
	\caption{Same as the left panel of Figure~\ref{fig:tE_piEparallel}, except for $\bar{v}_{\rm kick} = 200\,\kms$ (left) and $\bar{v}_{\rm kick} = 400\,\kms$ (right), respectively.}
	\label{fig:kick}
\end{figure*}

For natal kicks with $\bar{v}_{\rm kick} = 100\,\kms$, the selection efficiency remains high, around $90\%$, compared to the zero-kick case. However, for simulations with very high kicks, the efficiency drops significantly, to $\sim 50-60\%$ for $\bar{v}_{\rm kick} = 200\,\kms$ and only $\sim 10-20\%$ for $\bar{v}_{\rm kick} = 400\,\kms$. This decline reflects both reduced BH event rates and increased difficulties in filtering out contaminants with the selection criteria (see Figure~\ref{fig:kick}).

\section{Summary}

With Galactic simulations, we demonstrate that the $\pi_{\E,\parallel}/t_{\rm E}$ criterion is effective in selecting BHs with natal kicks $\lesssim 100$\,km/s for industrial-scale VLTI/GRAVITY+ mass measurements. We expect to detect about a dozen BHs per year by following up $\sim 35$ selected events with VLTI/GRAVITY+, making it promising to build a large sample within a few years to study the mass, distance, and transverse velocity distributions of isolated BHs in the Galactic bulge and disk.

\section*{Acknowledgment}
This work is supported by the National Natural Science Foundation of China (Grant No. 12133005) and the China Manned Space Project with No. CMS-CSST-2025-A16. SD acknowledges the New Cornerstone Science Foundation through the XPLORER PRIZE. This research was funded in part by National Science Centre, Poland, grants OPUS 2021/41/B/ST9/00252 and SONATA 2023/51/D/ST9/00187 awarded to PM.

\bibliographystyle{aasjournal}
\bibliography{refs}{}
\end{document}